\documentclass[twocolumn,prl,aps,showpacs]{revtex4}
\usepackage{graphicx}
\usepackage{doi}
\usepackage{hyperref}
\def\c60{C$_{60}$}

\def\d2I/dV2{d$^I/dV2$}

\begin{document}
\title{Ferromagnetic coupling of mononuclear Fe centers in a self-assembled metal-organic network on Au(111)}
\author{T. R. Umbach$^1$, M. Bernien$^1$, C. F. Hermanns$^1$, A.
Kr\"uger$^1$, V. Sessi$^2$, I. Fernandez-Torrente$^1$, P. Stoll$^1$$^,$$^3$, J. I.
Pascual$^1$$^,$$^3$$^,$$^4$, K. J. Franke$^1$$^,$$^3$$^,$$^5$, W. Kuch$^1$$^,$$^3$}
\affiliation{$^1$ Freie Universit\"{a}t Berlin, Fachbereich Physik, Arnimallee
14, 14195 Berlin, Germany \\
$^2$ European Synchrotron Radiation Facility, PB 220, 38043 Grenoble, France\\
$^3$ Center for Supramolecular Interactions, Freie Universit\"{a}t
Berlin, Arnimallee 14, 14195 Berlin, Germany\\
$^4$ CIC nanoGUNE, 20018 Donostia-San Sebastian, and ikerbasque, Basque Foundation for Science, 48011 Bilbao, Spain\\
$^5$ Institut f\"ur Festk\"orperphysik, Technische Universit\"at
Berlin, Hardenbergstra\ss e 36, 10623 Berlin, Germany}
\date{\today}

\begin{abstract}
The magnetic state and magnetic coupling of individual atoms in nanoscale structures relies on a delicate balance between different interactions with the atomic-scale surrounding. Using scanning tunneling microscopy, we resolve the self-assembled formation of highly ordered bilayer structures of Fe atoms and organic linker molecules (T4PT) when deposited on a Au(111) surface. The Fe atoms are encaged in a three-dimensional coordination motif by three T4PT molecules in the surface plane and an additional T4PT unit on top. Within this crystal field, the Fe atoms retain a magnetic ground state with easy-axis anisotropy, as evidenced by X-ray absorption spectroscopy and X-ray magnetic circular dichroism. The magnetization curves reveal the existence of ferromagnetic coupling between the Fe centers.
\end{abstract}


\maketitle 

As the size of a magnetic material decreases, the stability of the collective magnetic groundstate becomes weaker, leading to a paramagnetic behaviour at atomic scale dimensions \cite{Loth2012}. The identification and study of mechanisms for stabilizing magnetically oriented ground states aims to provide routes for the design and construction of nanomaterials (clusters and thin films) that retain some magnetic order at temperatures as high as possible. At the limit of individual atoms deposited on a metal surface, magnetic anisotropy and interatomic coupling are the key ingredients to tune the magnetism of a nanostructure \cite{Gambardella2009}. The anisotropy of the atomic environment may lower the magnetic moment but, in turn, defines easy directions for magnetization. Exchange interactions coupling the magnetic moment of neighbor atoms drive the formation of  magnetically ordered states, but this effect quickly vanishes with increasing interatomic distance. Only the indirect exchange coupling through valence electrons of the metal support (the Ruderman-Kittel-Kasuya-Yosida (RKKY) interaction \cite{Ruderman, Kasuya, Yosida,Wahl2007,Meier2008}) may be operative. It has been shown that RKKY interactions lead to collective ferromagnetic (FM) or anti-ferromagnetic (AFM) ground states, but the exchange energy is small ($J\sim 50$~$\mu$eV) \cite{Meier2008, Loth2012, Khajetoorians2012}.  

An interesting approach to enhance the magnetic anisotropy of atoms in a thin film is to augment the ligand field splitting of the $d$-levels by coordination to specifically selected organic ligands \cite{Gambardella2009}. The formation of a coordination bond to molecular endgroups induces a charge redistribution at the metal sites and cast the ligand field by adopting high-symmetry metal-organic structures and periodic networks \cite{Schlickum2007, Barth2009, Henningsen2011}. In bulk metal-organic materials, the organic linkers connecting magnetic metal centers also mediate the magnetic coupling between them by means of electronic states extended along the organic back-bone \cite{McCleverty1998, Wernsdorfer2002}. The presence of such superexchange mechanism for magnetic coupling in metal-organic thin films could optimize the stabilization of collective magnetic ground states, but its detection remains still a challenge. 

\begin{figure}
\begin{center}
\includegraphics[width=0.75\linewidth]{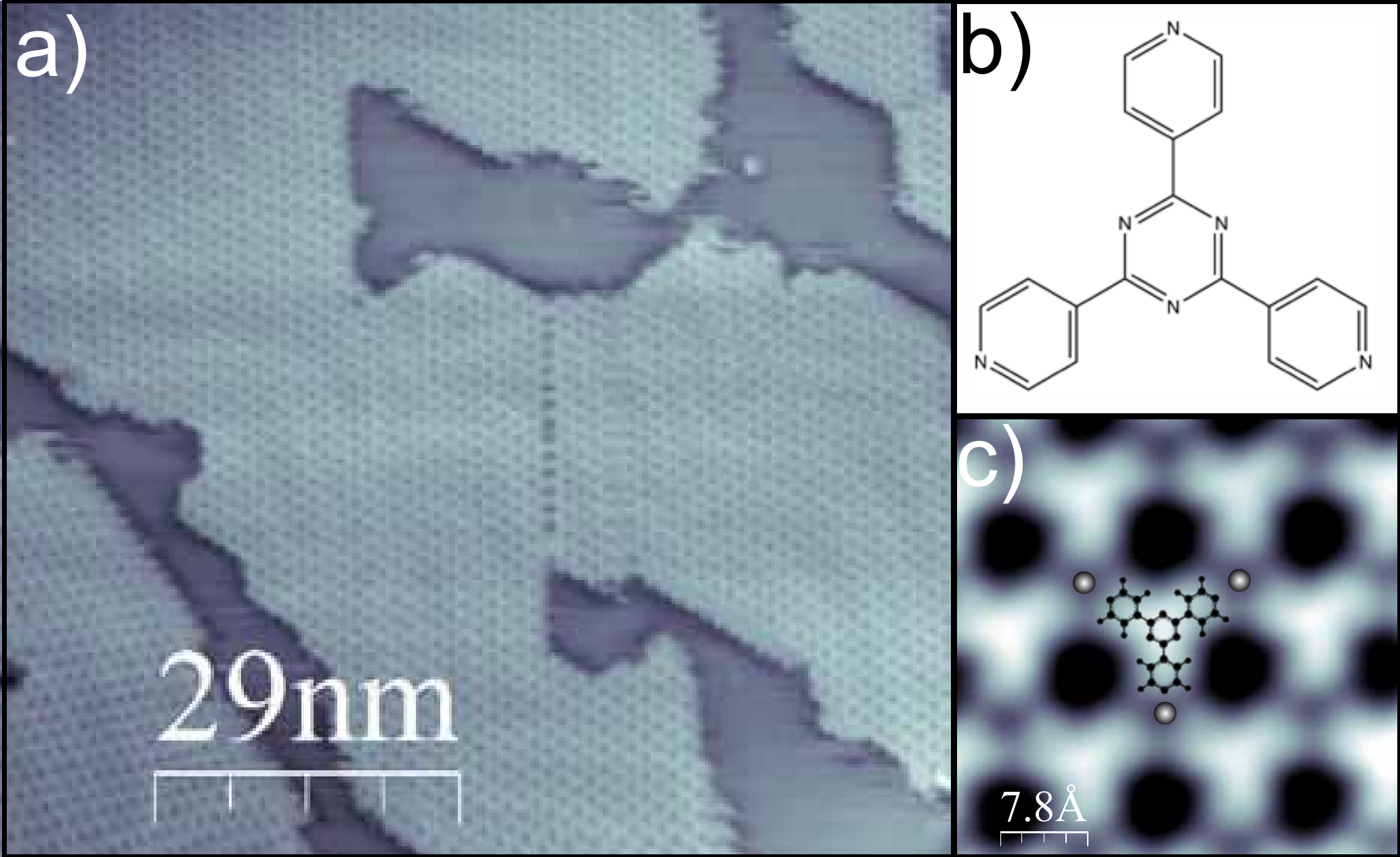}
\caption{\label{Fig1} (a) Large scale STM topography image of the
coordinated Fe-T4PT islands on the Au(111) surface ($I$=~253~pA, $V_{S}=-1.5$~V). The Fe-T4PT islands are homogeneous over the whole surface. There is no indication of Fe clusters. The STM image was obtained with a room temperature STM at the ID08 beamline at the European Synchrotron Radiation Facility (ESRF) directly before X-ray investigations. (b) Molecular structure of 2,4,6-Tris(4-pyridyl)-1,3,5-triazine (T4PT). (c) An STM zoom evidences the metal coordination of the T4PT molecules ($I$=~230~pA, $V_{S}$~=~0.06~V). STM image processing by WSxM \cite{Horcas2007}.}
\end{center}
\end{figure}

Here, we show the existence of ferromagnetic coupling between Fe ions within a self-assembled metal-organic film grown on a Au(111) surface. Single Fe atoms lie in three-dimensional coordination sites and are connected to neighboring Fe atoms by 2,4,6-tris(4-pyridyl)-1,3,5-triazine (T4PT) molecules (see Fig.~\ref{Fig1}b). Analysis of X-ray absorption spectroscopy (XAS) and X-ray magnetic circular dichroism (XMCD) measurements reveals a high spin $S$=2 ground state with an out-of-plane magnetic anisotropy. Magnetization curves of the ultra-thin layer exhibit deviations from the case of non-interacting paramagnetic moments, which are interpreted as a finite ferromagnetic interaction between the mononuclear centers on the basis of a mean-field model.

Fe-T4PT networks were grown on an atomically clean Au(111) single crystal. The gold surface was first precovered at room temperature by a submonolayer amount of T4PT molecules, sublimated from a Knudsen cell, and by Fe atoms. Subsequently, the substrate was annealed to 400~K to activate the formation of coordination bonds. The structure of the resulting surface was characterized in situ, in our low temperature (4.8~K) scanning tunneling microscope (STM). XAS and XMCD measurements in total electron yield were performed at the beamline ID08 at the European Synchrotron Radiation Facility (ESRF). The beamsize on the sample was about 1.4~mm (FWHM) in the horizontal, and 50~$\mu$m in the vertical direction. The sample was kept at a temperature of 8~K in a variable external magnetic field up to 5~T and exposed to X-rays under different incidence angles. Here, STM measurements (Fig.~\ref{Fig1}) were carried out in-situ at several macroscopically separated areas, indicating that the sample was covered by a homogeneous metal-organic network, and that there are no uncoordinated Fe clusters.

Large-scale STM images show the tendency of Fe atoms and T4PT molecules to coordinate in networks with high a degree of order (Fig.~\ref{Fig1}a). The presence of the Au(111) herringbone reconstruction underneath the metal-organic layer indicates a weak interaction with the substrate. STM images recorded at low bias voltage ($V_{S} <$~0.5~V) resolve the skeleton of the T4PT molecules (Fig.~\ref{Fig1}c). Each pyridyl group participates in a threefold node, a configuration that can only be stable if the natural repulsion between electrophilic endgroups is overcome by insertion of a bonding center prone to coordinate with the N lone-pair electrons \cite{Schlickum2007, Henningsen2011}. Hence, we anticipate the inclusion of Fe atoms at these sites \cite{Fe-note}.

\begin{figure}
\begin{center}
\includegraphics[width=0.95\linewidth]{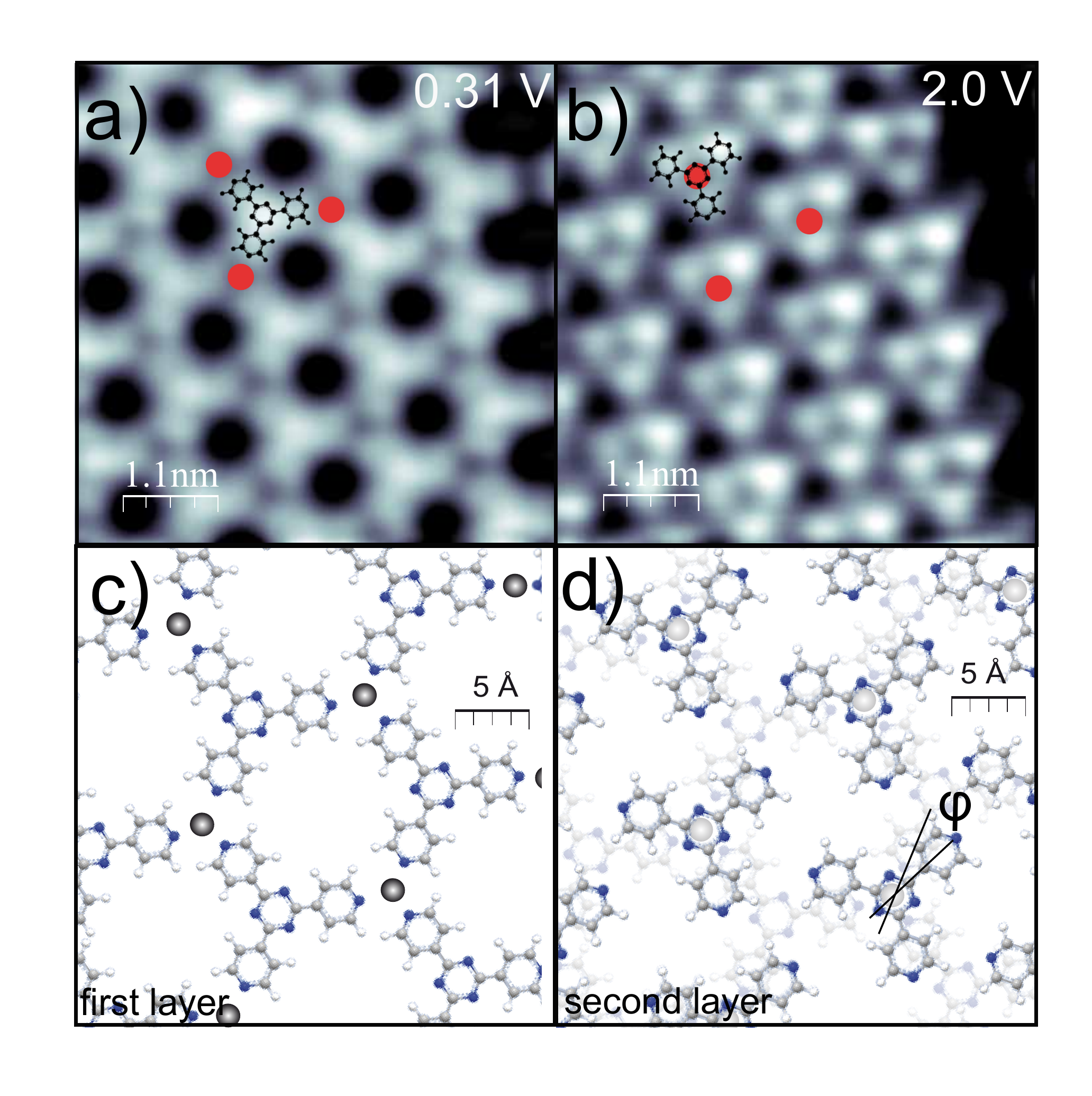}
\caption{\label{Fig2} (a)-(b) STM images of a Fe-T4PT network measured at different sample bias voltages  ($I$=~0.33~nA, (a) $V_{S}$~=~0.31~V, (b) $V_{S}$~=~2.0~V). (c) Structure of the first-layer of the Fe-T4PT network. Fe atoms are three-fold coordinated by three T4PT molecules in the first layer. (d) In the second layer, T4PT molecules are centered at the Fe sites and are rotated by $\varphi=29^{\circ}$ with respect to the first-layer molecules (indicated by the black lines).}
\end{center}
\end{figure}

The structure of the Fe-T4PT network shows a peculiar dependence on the bias applied during its inspection (Fig.~\ref{Fig2}a and b). Starting from the low bias structure in Fig.~\ref{Fig2}a, an increase of the sample bias above 0.6~V results in the observation of a new set of T4PT molecules centered on the Fe sites and rotated by $\varphi=29^{\circ}$ (Fig.~\ref{Fig2}b). These new features can not be associated to tunneling through different orbitals of the same molecules due to their different orientation. Furthermore, we observe domains with mirror symmetry around dislocation lines at high bias voltage that are absent when inspected at low bias voltage (see supplementary information S1). Based on these facts, we conclude that the film consists of a bilayer of T4PT molecules: Fe atoms coordinate to the pyridyl groups of three T4PT in the bottom layer (first layer in Fig.~\ref{Fig2}c) and to the triazine center of an additional T4PT molecule directly on top (second layer in Fig.~\ref{Fig2}d). Their different aspect with the bias voltage can be rationalized by considering that the top layer is only weakly coupled to the metallic substrate and poorly screened \cite{Torrente2008}. Its electronic gap becomes wider and the molecule more transparent to electrons tunneling close to the Fermi level ${E_F}$ \cite{Repp2005,Takada2004}, allowing us to resolve the structure of the bottom layer at low bias voltages.

In the resulting structure the Fe atom is located in a three-dimensional coordination cavity, encaged by a total of six nitrogen atoms (Fig.~\ref{Fig2}c and d). The presence of the top T4PT centered directly on Fe sites suggests that the triazine moiety also participates in the coordination with the metal atom.  A probable scenario is that this metal-ligand bond causes an upward displacement of the Fe atom \cite{Flechtner2007, Henningsen2011,Isvoranu2011}, which would cause a distortion of the ligand field around the Fe atom, bringing it from a trigonal-planar into a three-dimensional (pyramidal) structure \cite{Li2008,Halcrow2007}. This configuration could help to prevent the quenching of its magnetic moment due to the interaction with the gold surface.

\begin{figure}
\begin{center}
\includegraphics[width=0.9\linewidth]{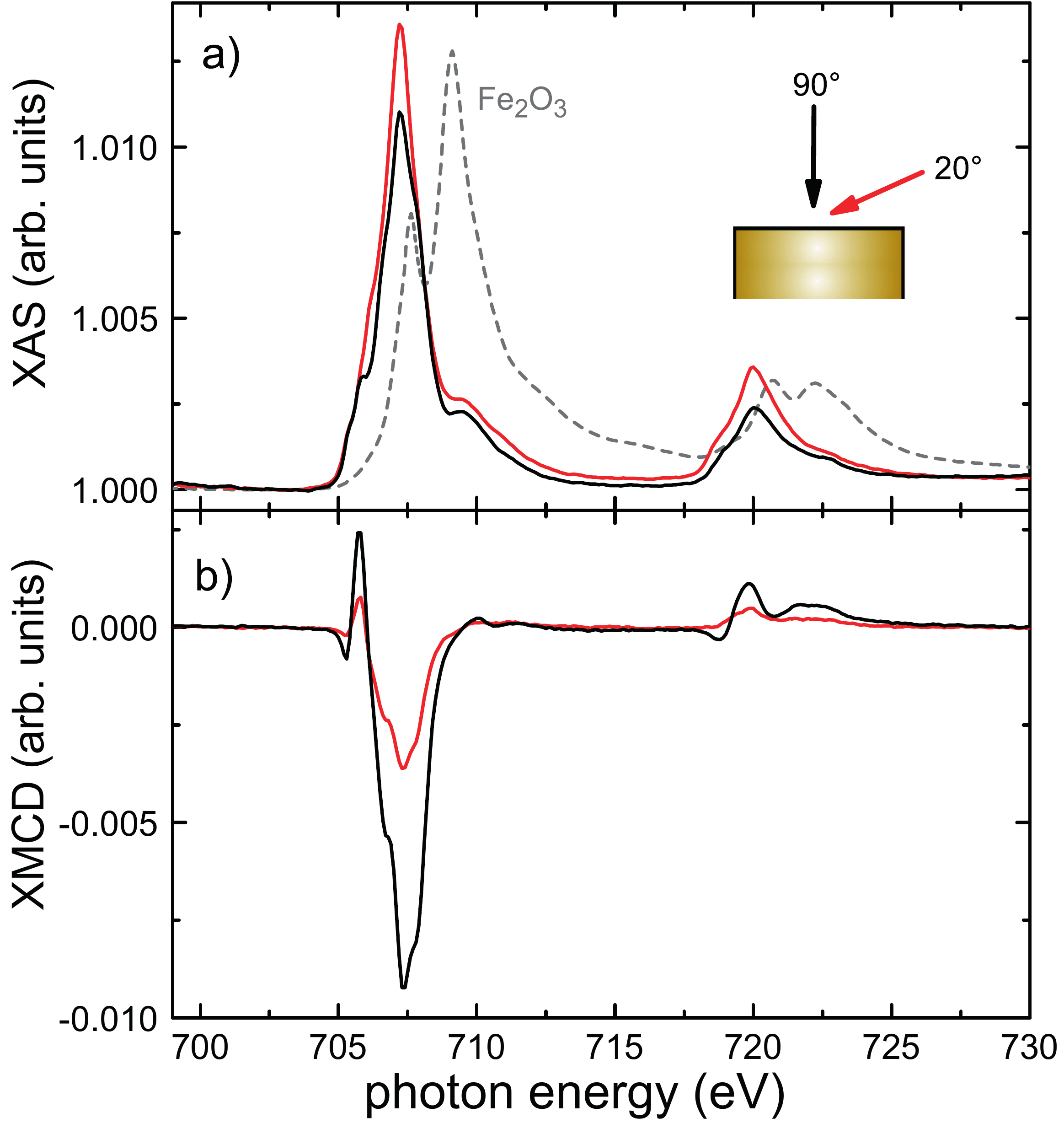}
\caption{\label{Fig3} (a) X-ray absorption spectra and (b) XMCD difference spectra at the Fe $L_{2,3}$ edges of Fe-T4PT on Au(111), measured
at a temperature of 8~K in an applied magnetic field of 5~T for two different angles of incidence, as defined in the sketch in the top panel.
The magnetic field was always parallel to the X-ray beam. The dashed line represents an XAS of the $L_{2,3}$-edge in Fe$_2$O$_3$, where Fe is in a +3 oxidation state. The shift of the absorption edge is due to the chemical sensitivity of the binding energy of the core-level electrons.}
\end{center}
\end{figure}

To investigate the magnetic properties of the Fe atoms in the Fe-T4PT networks, we have performed XAS and XMCD measurements at $T$~=~8~K in an applied magnetic field of 5~T (Fig.~\ref{Fig3}). From the line shape and position of the XAS $L_3$ peak (Fig.~\ref{Fig3}a, maximum at 707.2~eV) we conclude that we probe isolated Fe atoms  in a +2 oxidation state \cite{noteatoms, Laan1992, Zheng2005, Regan2001} (the spectrum clearly deviates from that of Fe(III), shown with a dashed line in Fig.~3a for comparison). A +2 oxidation state is typical in bulk mononuclear iron complexes \cite{Halcrow2007}, thus supporting the three-dimensional coordination motif described in Fig.~\ref{Fig2}d. Multiplet calculations of the XAS and XMCD difference spectra indicate an $S=2$ high-spin state (see supplementary information S2).

The presence of a magnetic moment is further corroborated by XMCD spectra shown in Fig.~3b. The XMCD difference signal of the $L_3$ peak exhibits a characteristic dip-peak structure at the low-energy side. While the line shapes are similar for different angles of incidence, there is a large change in the XMCD peak height: the signal at normal incidence is more than a factor of two higher than at grazing incidence ($20^{\circ}$). The higher XMCD signal in normal incidence points towards an easy-axis magnetic anisotropy along the direction normal to the surface, which is a result of the spin-orbit-induced mixing of nearly degenerate Fe $3d$ levels in the surrounding ligand field.

\begin{figure}
\begin{center}
\includegraphics[width=0.9\linewidth]{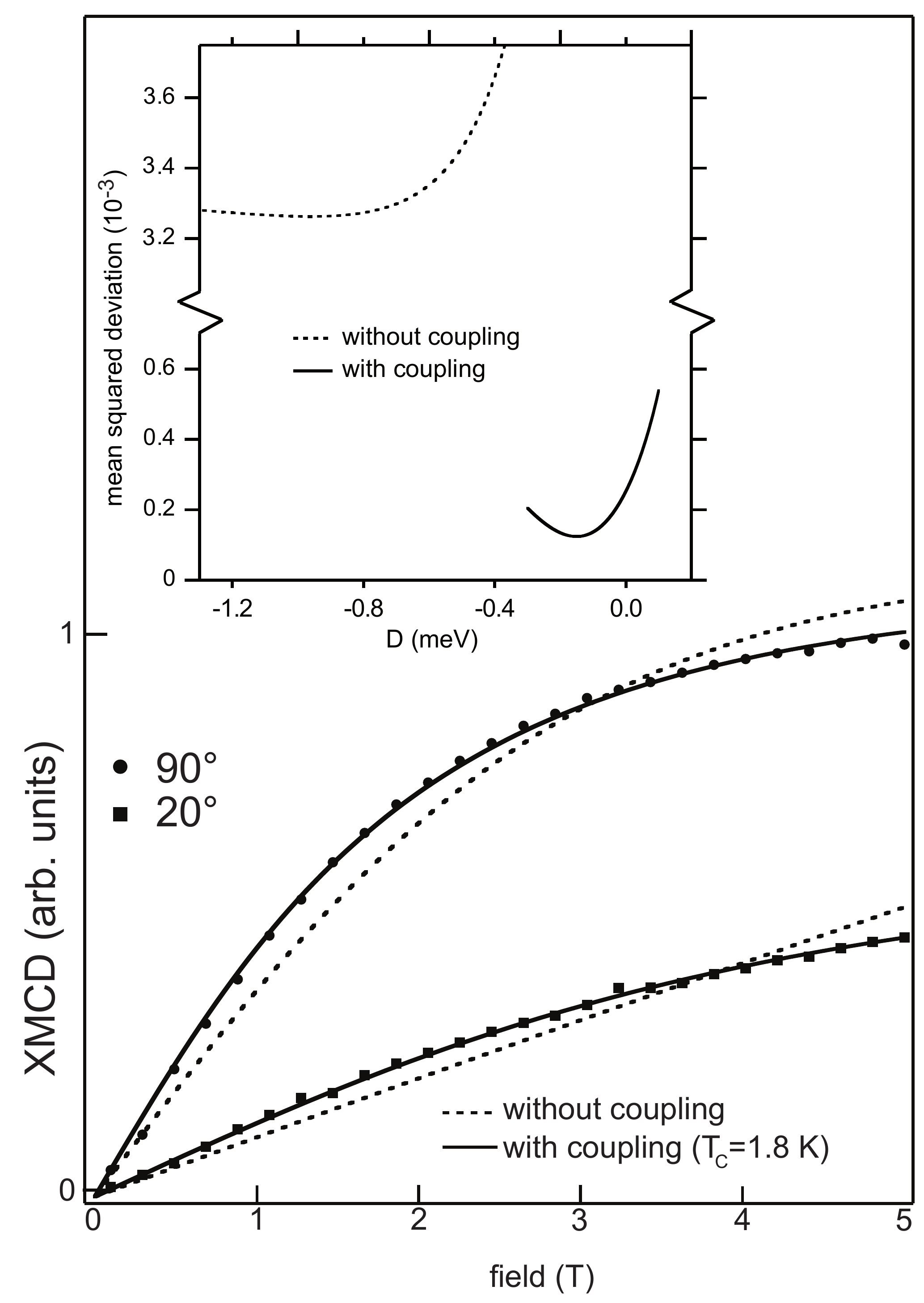}
\caption{\label{Fig4} Magnetic field dependence of the XMCD at the Fe $L_3$ edge of Fe-T4PT on Au(111) for X-ray incidence normal to the surface (circles) and at $20^{\circ}$ grazing incidence (squares), measured at a temperature of 8~K. The magnetic field was always parallel to the X-ray beam. Dashed lines are the best fits of a
paramagnetic spin Hamiltonian for $S$=2, $T$=8 K, and $D=-1.16$~meV, solid lines are the result of a simulation using mean-field-coupled magnetic moments with $T_C$ = 1.8 K (see text). The inset shows the mean squared deviation between the experimental field dependence of XMCD data and the simulation by the paramagnetic spin Hamiltonian (dashed line) and the mean-field model, as described in the main text, as a function of the anisotropy, represented by the zero-field splitting parameter $D$. For each value of $D$, all other fit parameters have been optimized.}
\end{center}
\end{figure}

Fig.~\ref{Fig4} shows the evolution of the Fe $L_3$ XMCD signal with applied external magnetic field for both perpendicular and grazing incidence. The curvature of these plots allows us to quantify the anisotropy of the system. 
We start first by simulating the magnetization curves assuming non-interacting paramagnetic moments under the presence of magnetic anisotropy by means of a spin Hamiltonian approach:\\
\begin{eqnarray}
\mathcal{H} = D S_{z}^{2} - \mu_B g \bf{B} \cdot \bf{S}
\label{spinH}
\end{eqnarray}

The first term describes the anisotropy energy through the zero-field-splitting parameter $D$ and $S_z$, the spin component normal to the molecular plane (a negative $D$ stands for an easy axis along z). The second term accounts for the Zeeman energy ($\bf{S}$ is the spin vector, $\bf{B}$ the vector of the external field, $\mu_B$ Bohr's magneton, and we consider a $g$ factor of 2). Solving Eq.~\ref{spinH} for the two directions of measurement and calculating the magnetization from the thermal population of the resulting levels allows us to simulate magnetization plots like in Fig.~\ref{Fig4} for several conditions. The curvature of the magnetization plots is an expression of the anisotropy of the system and can be tuned by adjusting the parameter $D$. The larger curvature in normal incidence corroborates that this direction is the easy axis for magnetization ($D<0$). However, the experimental curvature cannot be properly reproduced, no matter what value of $D$ is used. The dashed line in the inset of Fig.~\ref{Fig4} shows the mean squared deviation between experiment and simulations evaluating Eq.~\ref{spinH} as a function of $D$. We obtain large standard deviations for all values of $D$, even when the curvature is simulated using unrealistically high $D$ values. We thus conclude that the experimental data are not compatible with non-interacting anisotropic paramagnetic moments, but bear instead evidence for a magnetic coupling between neighboring Fe moments \cite{noteFeCluster}.

To include magnetic interactions between nearest neighbors, we resort to a mean-field description to simulate the magnetization curves. In this model, the magnetic interaction of a spin with all other spins in the neighborhood is replaced by an effective mean magnetic field added to the external applied field and proportional to the magnetization of the ensemble with proportionality factor $T_C$=$\sum\limits_{i}$ $J_{i}$/(3$k_B$). Here $J_i$ is a Heisenberg-type interaction of the spin with a spin at site $i$, $k_B$ is the Boltzmann constant. The sum goes over all sites with non-vanishing $J_i$. $T_C$ represents the Curie temperature in the limit of vanishing anisotropy. 

The solid lines in Fig.~\ref{Fig4} are the result of such a simulation with $D=-0.35$~ meV and $T_C=1.8$~K. These values correspond to a global minimum (see solid line in the inset of Fig.~4 and supplementary information S4), reproducing correctly the experimental magnetization curves. While the model may not capture the details of the coupling mechanism, it clearly evidences the necessity to include ferromagnetic coupling of the Fe atoms to explain the experimental curves. If we assume a coupling to only the six nearest neighbor Fe atoms, the coupling constant $J_{i}~=~k_B~T_C/2$  has to be of the order $\approx$~80~$\mu$eV to reproduce the experimentally observed magnetization curves \cite{noteSpin1}. Since the measurements are carried out well above $T_C$, the effect of correlated fluctuations, which would lead to an underestimation of the derived coupling strength in the mean-field description, are small.

Due to the relatively large distance between the Fe centers ($r=1.3$~nm) a direct exchange coupling is very unlikely to be the origin of the observed ferromagnetic coupling. An indirect coupling mechanism via the polarization of the surface and bulk conduction electrons has been shown to exist on metal surfaces \cite{Wahl2007, Meier2008, Zhou2010, Tsukahara2011}. The RKKY-interaction exhibits an oscillating
ferro-/anti-ferromagnetic coupling with the distance $r$ of the magnetic moments. It can be described as $J_{RKKY} \propto
\frac{cos(2k_{F}r)}{(2k_{F}r)^d}$ \cite{Fischer1975, Meier2008}, where $k_{F}$ is the Fermi wave vector, and $d$ defines the dimensionality of the conduction electron system. The Fermi wave vector of the Rashba-split surface state amounts to $k_{F,1}\approx$~1.7~nm \cite{Reinert2001}. Hence, a pairwise interaction of the Fe atoms at a distance of 1.3 nm via RKKY interactions would result in a weakly coupled anti-ferromagnetic ground state. Furthermore, we expect that the sixfold ligated Fe atoms are lifted from the surface \cite{Flechtner2007, Isvoranu2011} and thus only interact weakly with the substrate electrons. Despite small effects that may change the scattering phase shift of the conduction electrons, we may tentatively exclude a significant contribution of an RKKY-mediated coupling. 

The molecular ligands coordinated to two neighbor Fe atoms can also play an important role in mediating magnetic interactions through a superexchange mechanism. Ferromagnetic and anti-ferromagnetic exchange interactions in metal-organic frameworks are well known in bulk materials \cite{McCleverty1998} for nitrogen-based organic linkers and transition metal atoms. The superexchange mechanism is enhanced when conjugated molecular states bridge the magnetic centers \cite{Bellini2011}, as it is the case for Fe-T4PT networks. The spin polarization qualitatively follows a simple alternation rule: an even number of linker atoms in the pathway leads to anti-ferromagnetic coupling, while an odd number results in ferromagnetic coupling. In the case of the Fe-T4PT network, the alternation rule results in a ferromagnetic coupling between neighboring Fe centers. This is consistent with the positive (ferromagnetic) Heisenberg-type interaction
$J_{i}$ deduced from the magnetization curves.

Our results clearly sustain the presence of ferromagnetic coupling of Fe atoms connected by organic molecules in a highly ordered metal-organic network. The peculiarity of the observed network is that it consists of a bilayer structure, in which the Fe is coordinated by three pyridin and one triazine unit. The central triazine ligand may further be responsible for a lifting of the Fe atoms from the metallic surface due to the trans-effect \cite{Flechtner2007, Isvoranu2011}. This configuration eases the formation of a +2 oxidation state with a net spin. The resulting magnetic moments of the Fe ions are coupled ferromagnetically with each other, probably via a superexchange mechanism mediated by the organic linkers. These results further demonstrate that magnetism in two-dimensional metal-organic networks on a metal surface is feasible without the use of ferromagnetic substrates and suggests a route to engineer its magnetic character by enclosing the magnetic centers in self-assembled three-dimensional coordination cavities. To increase the stability of such magnetic networks for potential applications, a larger anisotropy and significantly stronger magnetic coupling strengths are required. Chemical design strategies towards this goal may comprise the use of radical ligands, known from bulk metal-organic ferromagnetic materials \cite{InoueJACS94, InoueJACS96}.

We thank Hendrik Mohrmann for fruitful discussions. We also acknowledge financial support by the Deutsche Forschungsgemeinschaft (grant FR 2726/1 and Sfb 658) and by the Center
for Supramolecular Interactions (CSI) of Freie Universit\"at Berlin.

\newpage

\begin{center}
\bfseries{\Huge{Supplemental Material}}
\end{center}
%
\quad 
\quad 
\begin{center}
\Large{T. R. Umbach$^1$, M. Bernien$^1$, C. F. Hermanns$^1$, A. Kr\"uger$^1$, \\
V. Sessi$^2$, I. Fernandez-Torrente$^1$, P. Stoll$^{1,3}$, J. I. Pascual$^{1,3,4}$, \\
K. J. Franke$^{1,3,5}$, W. Kuch$^{1,3}$}
\end{center}
\quad 
\quad 
\begin{center}
$^1$ Freie Universit\"{a}t Berlin, Fachbereich Physik, Arnimallee 14, 14195 Berlin, Germany

$^2$ European Synchrotron Radiation Facility, PB 220, 38043 Grenoble, France

$^3$ Center for Supramolecular Interactions, Freie Universit\"{a}t Berlin, Arnimallee 14, \\ 14195 Berlin, Germany

$^4$ CIC nanoGUNE, 20018 Donostia-San Sebastian, and \\ ikerbasque, Basque Foundation for Science, 48011 Bilbao, Spain

$^5$ Institut f\"ur Festk\"orperphysik, Technische Universit\"at Berlin, Hardenbergstra\ss e 36, \\ 10623 Berlin, Germany

\end{center}


\section{STM bias dependence of the Fe-T4PT coordination network}
\begin{figure}
\begin{center}
\includegraphics[width=0.8\linewidth]{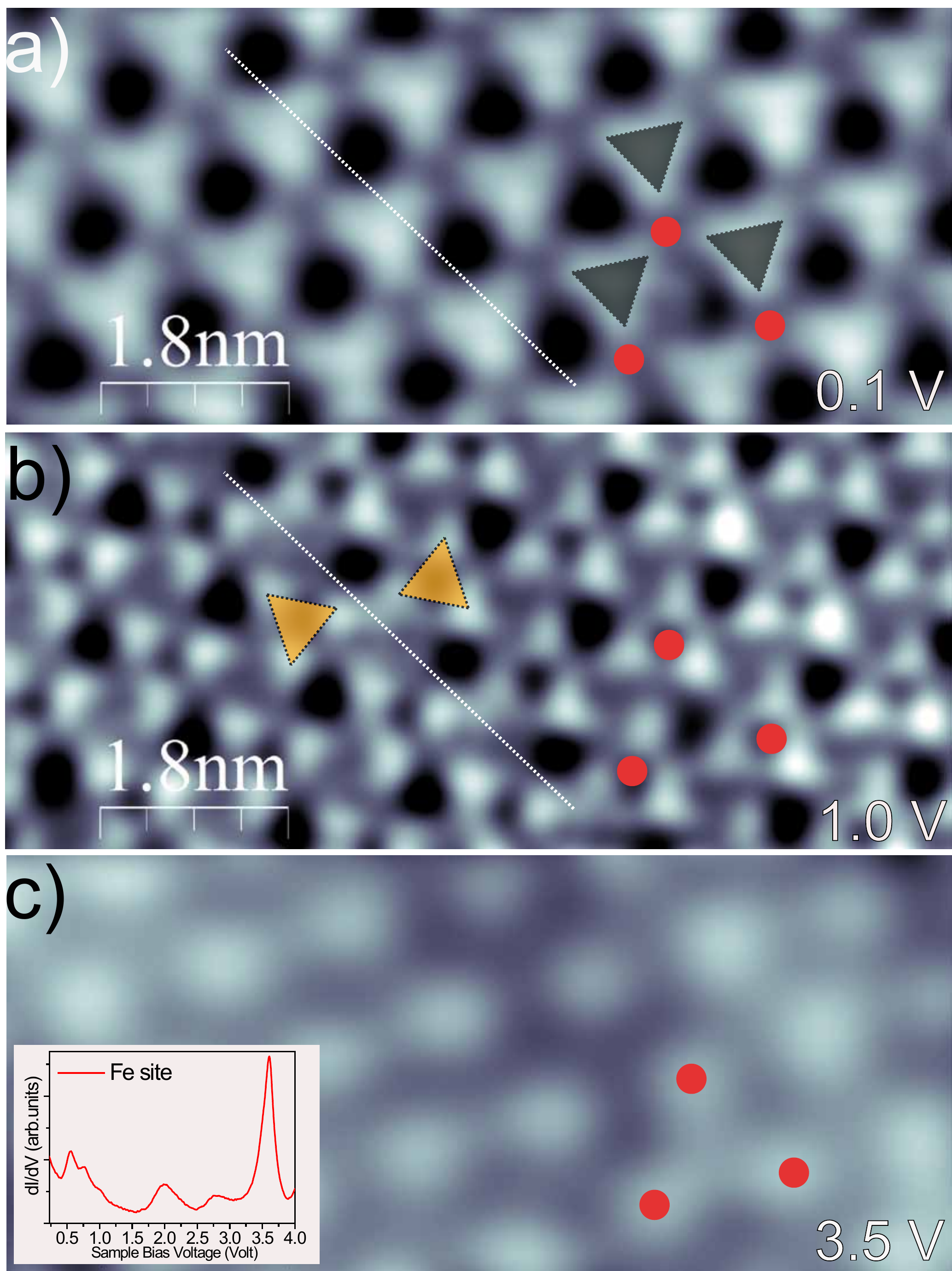}
\caption{\label{FigS6}(a-c) STM topography images of the Fe-T4PT network at different sample bias voltages $V_{S}$ (0.1~V, 1.0~V, and 3.5~V). At 0.1~V the first molecular layer is visible. At 1.0~V the second molecular layer is imaged. The white dashed lines indicate a dislocation line in the second layer, where the second-layer T4PT molecules change their orientation. This change cannot be observed in the first layer (see panel (a)). The inset in panel (c) shows a $dI/dV$ spectrum at an Fe center with resonances at 0.5~V, 2.0~V, 2.8~V, and 3.5~V. The resonance at 3.5~V can be associated with a localized electronic state of the Fe atom.}
\end{center}
\end{figure}

The STM images of the Fe-T4PT network show a strong dependence on the applied bias voltage (Figs.~\ref{FigS6} (a-c)). At low sample bias voltage (0.1~V), we resolve the shape of the T4PT molecules, where each pyridyl group points to a threefold node. This configuration is stabilized by the coordination to an Fe atom. At larger sample bias voltage (1.0~V in Fig.~\ref{FigS6} (b)) the appearance of the network changes drastically. We observe triangles formed by three distinct protrusions. The center of these triangular shapes are located on top of the Fe atoms and their rotational orientation does not match any high-symmetry of the molecular shapes found in Fig. \ref{FigS6}~(a). Furthermore, we note that the orientation of the triangles is not the same throughout the whole image (indicated by the white line), whereas this is the case for the molecules seen at low bias voltage. Hence, we interpret the features at 1~V as a second layer of molecules, which may adopt different orientations with respect to the underlying molecular layer. The protrusions represent the pyridyl moieties of the T4PT molecules. Increasing the sample bias voltage further to 3.5~V leads again to a significant change of the contrast (shown in Fig.~\ref{FigS6} (c)). Bright protrusions at the Fe sites of the Fe-T4PT network are observed. Their identification as Fe atoms is sustained by a pronounced resonance in the local differential conductance spectra at 3.5~V at these sites (inset in Fig.~\ref{FigS6} (c)), probably arising from the coordinated $d$-states \cite{Bjork2010, Henningsen2011}. The appearance and disappearance of intensity in the STM images can be understood by tunneling through localized molecular electronic states. In particular, the second layer is weakly coupled to the substrate, thus exhibiting a poor conductivity at low bias voltages. The peculiar bias dependence is a proof of the two-layer structure of the Fe-T4PT network: each Fe atom is surrounded by three pyridyl groups of three first-layer T4PT molecules and one triazine ring of a second-layer T4PT molecule.
%
\begin{figure}
\begin{center}
\includegraphics[width=0.8\linewidth]{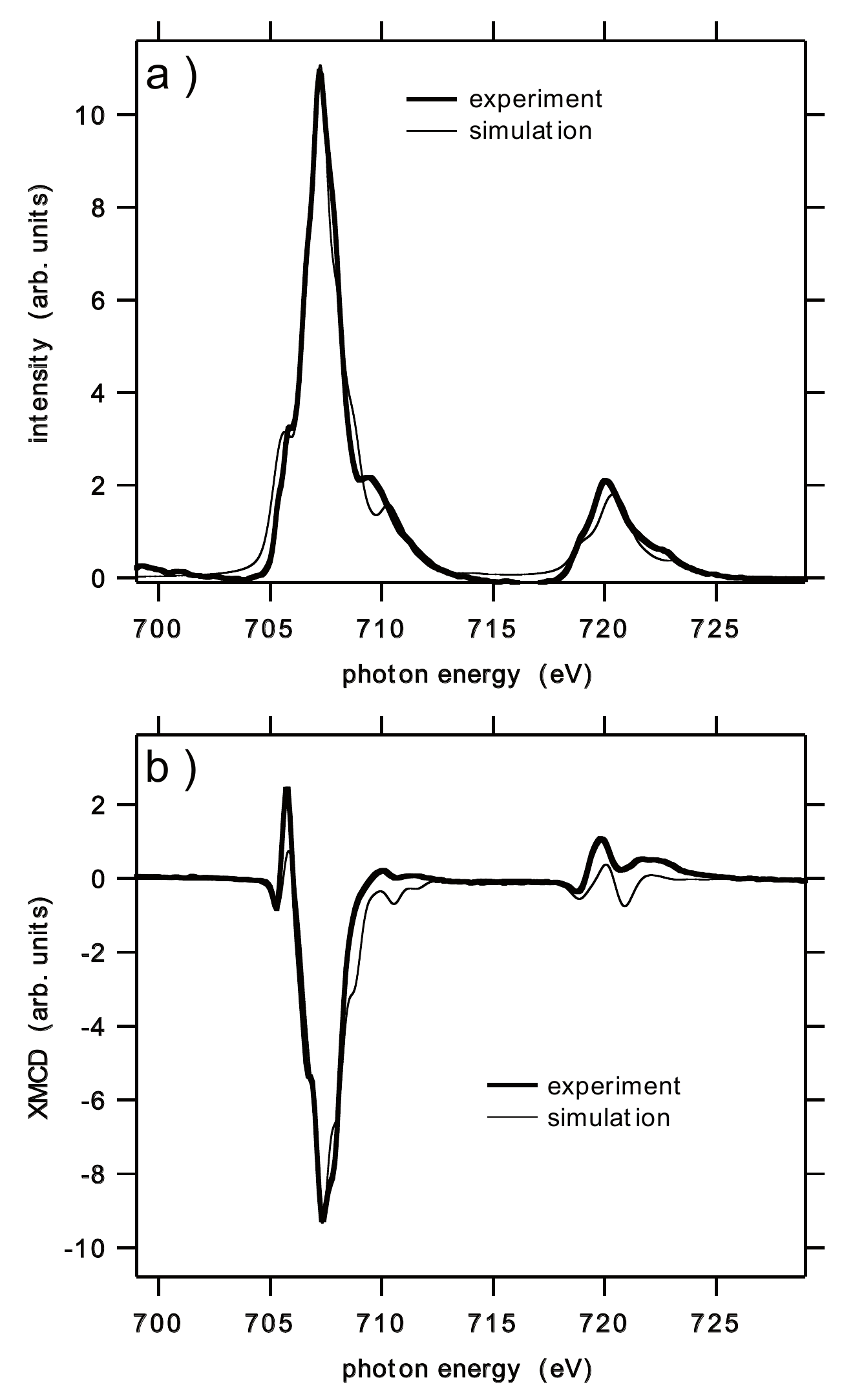}
\caption{ \label{S1} (a) Helicity-averaged x-ray absorption spectra and (b) XMCD difference curves at the Fe $L_{2,3}$ edges of Fe-T4PT on Au(111). Thick lines: Experimental spectra, measured at a temperature of 8 K in an applied magnetic field of 5 T in normal incidence of circularly polarized radiation. Thin lines: Result of a multiplet simulation, see text.}
\end{center}
\end{figure}

\section{Multiplet calculations}

The x-ray absorption spectrum and XMCD difference spectrum for normal incidence have been simulated by multiplet calculations using Cowan's code \cite{Cowan1981} and the CTM4XAS program version 5.5 \cite{Stavitski2010}. A good agreement for both, the spectral shape of the helicity-averaged absorption spectrum and the XMCD difference curve, is found for the following crystal field parameters: $10Dq=940$~meV, $D \tau=40$~meV, and $D \sigma =-200$~meV. For these parameters, which describe a trigonal ($D_{3d}$) crystal field, the Fe ion is in an $S=2$ high-spin state. Fig.~\ref{S1} shows the resulting simulations (thin lines) together with the experimental spectra (thick lines, as also shown in Fig.~3 of the main text) for the helicity-averaged x-ray absorption spectrum (a) and the corresponding XMCD difference curve (b).
\begin{figure}
\begin{center}
\includegraphics[width=0.8\linewidth]{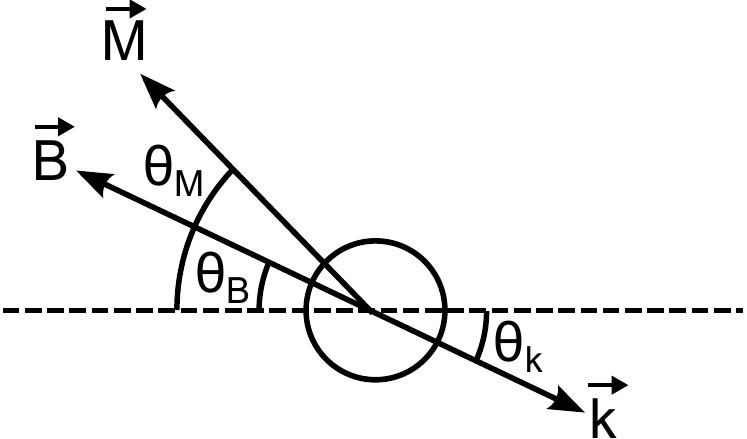}
\caption{\label{S7} Relative alignment of the magnetic field \textbf{B}, the magnetization \textbf{M}, and the \textbf{k} vector of the x-rays.}
\end{center}
\end{figure}

\begin{figure}
\begin{center}
\includegraphics[width=0.8\linewidth]{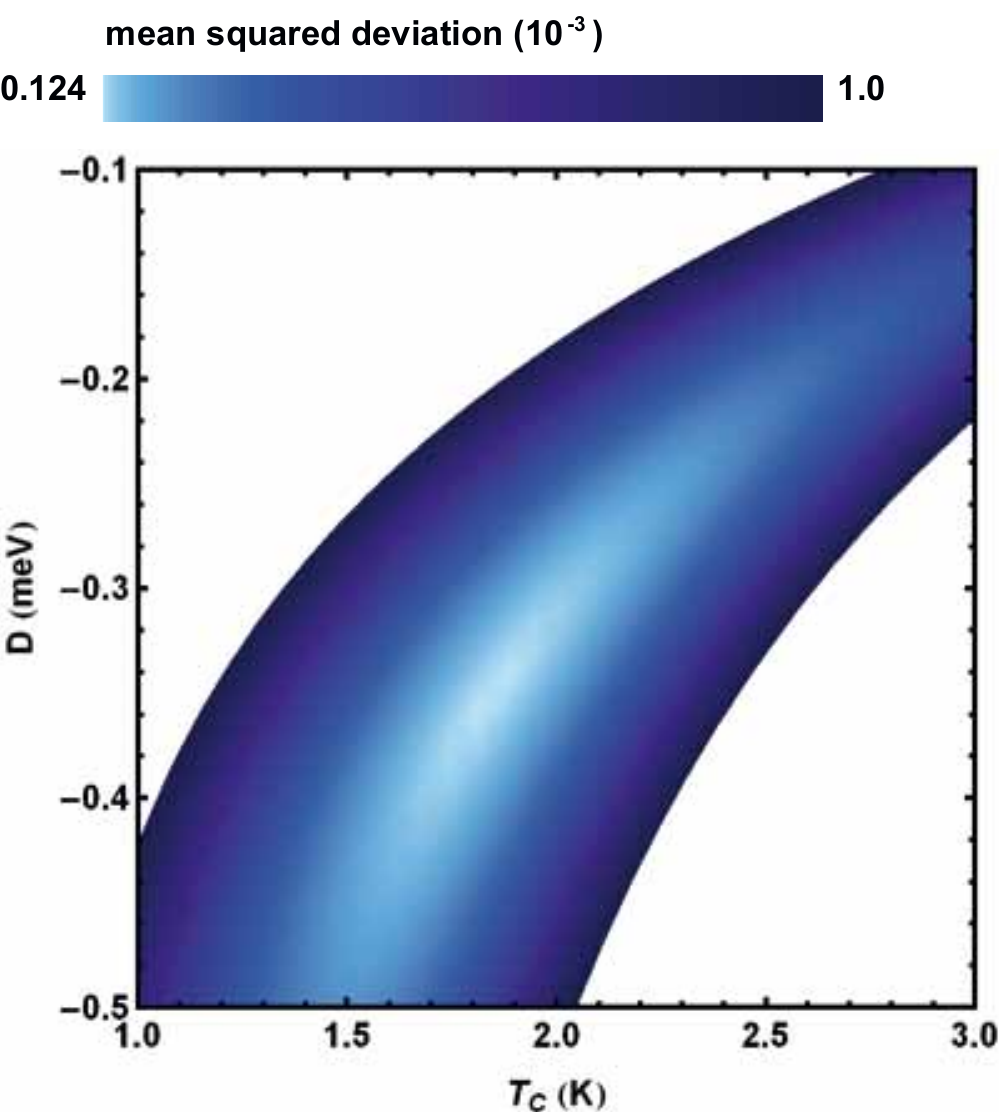}
\caption{\label{S2} Two-dimensional color plot of the mean squared deviation of the simulation calculated by the mean-field model described in the main text from the experimental field-dependence of XMCD data as a function of the magnetic anisotropy, represented by the zero-field splitting parameter $D$, and the magnetic coupling between the Fe ions, expressed by $T_{C}$. The deviation is calculated from the simultaneous comparison to the experimental XMCD values for normal and $20^{\circ}$ grazing incidence. The white regions outside the curved valley of minimal deviation exhibit deviation values larger than $10^{-3}$.}
\end{center}
\end{figure}

\section{Fit of the magnetization curves}

The field- and angle-dependent XMCD signals presented in Fig.~4 of the main text are not directly proportional to the magnetization \textbf{M} of the Fe ions that is modeled by the spin-Hamiltonian. If the magnetic field \textbf{B} is not applied along the easy axis of the ions, their magnetic moments are not aligned parallel with the magnetic field, and the XMCD signal detects only the projection of the magnetization onto the \textbf{k} vector of the x-rays. The relative alignment of the magnetic field, the magnetization, and the \textbf{k}  vector as well as the definition of the corresponding angles are given in Fig.~\ref{S7}. The XMCD signal depends further on the magnetic dipole operator and its angle dependence. This term accounts for the anisotropy of the spin-density distribution around the Fe ion and can be large for isolated ions. Its temperature dependence is the same as the one of the magnetization, and its angle dependence is given by $(1-3\cos(2\theta_M))$ if the spin-orbit coupling is small compared to the ligand-field splitting \cite{Stohr1995}. The intensity of the XMCD signal is calculated from the spin-Hamiltonian model according to:

\begin{center}
	$I_{XMCD}=A \cos(\theta_M-\theta_k )\left|\textbf{M}\right|(1-C(1-3 \cos(2\theta_M )))$
\end{center}

where $A$ is an overall scaling factor to the experimental signal and $C$ is a factor determining the relative contribution of the magnetic dipole operator. Besides $T_C$ and $D$, also $A$ and $C$ were treated as fit parameters for fitting the model to the experimental data. 



\section{Error margins}
The mean squared deviation between the experimental data points of the magnetic field dependence of the XMCD difference spectra at the Fe $L_{3}$ edge and the fit regarding to the mean-field model of coupled anisotropic magnetic moments is shown as a two-dimensional plot in Fig.~\ref{S2} as a function of the zero-field splitting parameter $D$ and the coupling strength, expressed by $T_{C}$. In addition, $C$ and $A$ have been fitted for each set of $D$ and $T_{C}$. From Fig.~\ref{S2} one can see that, although there is some correlation between $D$ and $T_{C}$, a clear minimum is found at the parameters mentioned in the main text, i.e., $D=-0.35$~meV and $T_{C}=1.8$~K.
\\
\\



\begin{figure}
\begin{center}
\includegraphics[width=0.8\linewidth]{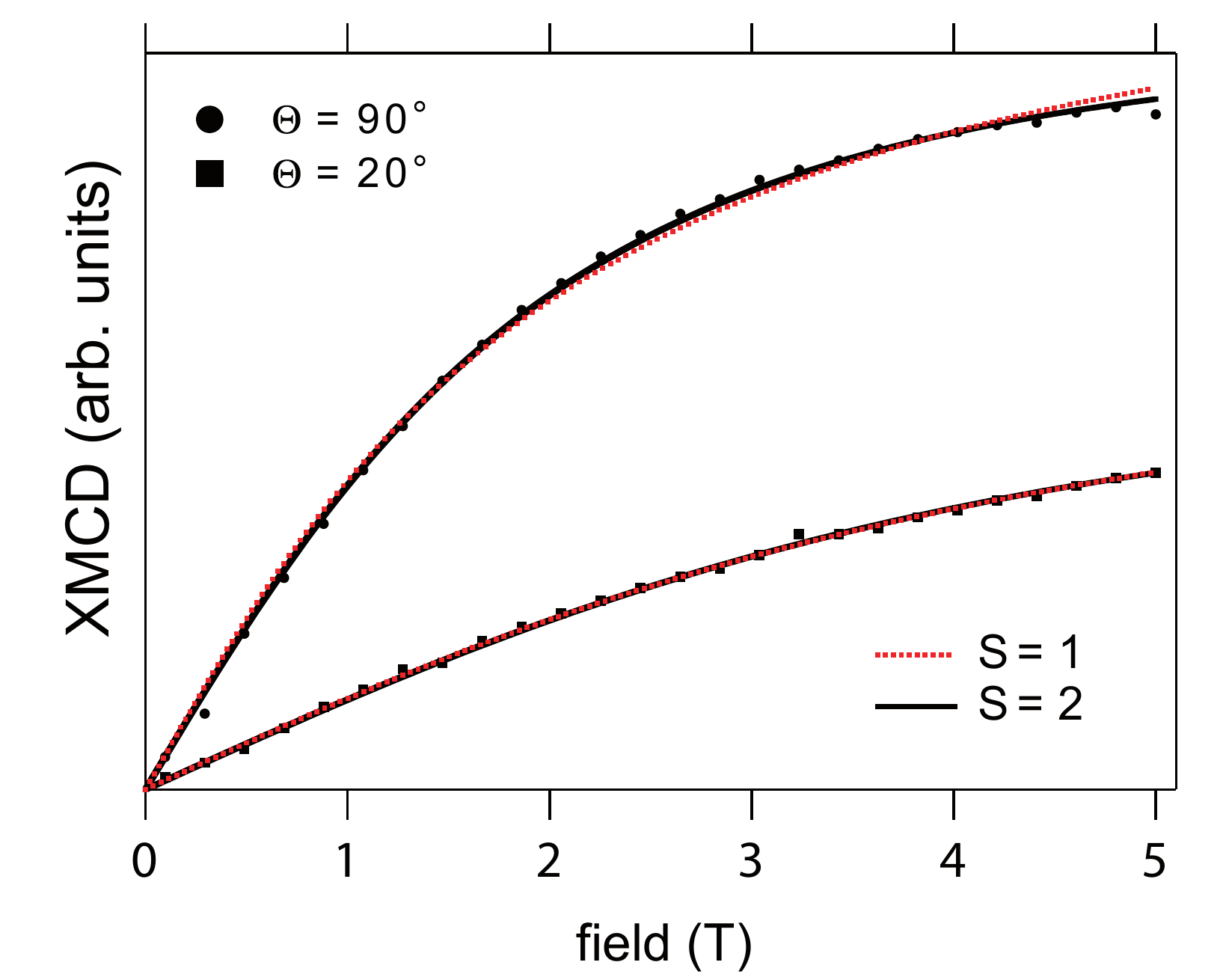}
\caption{\label{S4} Magnetic field dependence of the XMCD at the Fe $L_{3}$ edge of Fe-T4PT on Au(111) for x-ray incidence normal to the surface (circles) and at 20$^{\circ}$ grazing incidence (squares), measured at a temperature of 8 K. The solid black line is the best fit of mean-field-coupled magnetic moments with $S$=2, the red dotted line for moments with $S$=1.}
\end{center}
\end{figure}

\section{Spin state}

Fig.~\ref{S4} shows a comparison of the best fit of the mean-field model described in the main text for $S=1$ (red dashed lines) to the field-dependence of the experimental XMCD data (solid symbols) together with the best fit for $S$=2 (black lines). (The latter two are also shown in Fig.~4 of the main text.) Similarly good fits are obtained for $S=1$ and $S=2$. For $S=1$ a larger anisotropy and a larger coupling between Fe ions result from the fit, with values of	$D=-1.27$~meV and  $T_{C}=3.7$~K.

%
%
%


\begin{thebibliography}{33}
\bibitem{Fe-note} Growth of T4PT films without deposition of Fe atoms resulted in a completely different layer structure, where direct bonds between N atoms of pyridyl moieties are fully absent.
\bibitem{noteFeCluster} We rule out that the magnetization is due to Fe clusters as we do not observe their presence in any STM images taken at various sample locations of the very same preparation of the Fe-T4PT network. Uncoordinated Fe clusters can be further excluded by the lineshape \cite{noteatoms} of the XAS and XMCD signal, and by the magnetization behavior. The magnetization of potential residual Fe clusters would have an in-plane anisotropy \cite{OhresserPRB01}, contrary to our observed easy-axis anisotropy.
\bibitem{noteSpin1} This conclusion is robust and does not depend on the assumption of an $S=2$ high-spin state. Even stronger magnetic coupling between Fe atoms is necessary to reproduce the experimental data of Fig.~\ref{Fig4} if  $S=1$.
\end{thebibliography}
\end{document}